\documentclass[9pt,twocolumn,twoside]{pnas-new}

\templatetype{pnasresearcharticle} 

\title{Symmetric shear banding and swarming vortices in bacterial ``superfluids''}

\author[a,b]{Shuo Guo}
\author[a,1]{Devranjan Samanta} 
\author[a]{Yi Peng}
\author[b,2]{Xinliang Xu}
\author[a,2]{Xiang Cheng}

\affil[a]{Department of Chemical Engineering and Materials Science, University of Minnesota, Minneapolis, MN 55455, USA}
\affil[b]{Beijing Computational Science Research Center, Beijing 100193, China}

\leadauthor{Guo} 

\significancestatement{Bacterial suspensions can flow without apparent viscosity. Such a superfluid-like behavior stems from the collective motions of swimming bacteria. Here, we explore the microscopic flow profile of bacterial ``superfluids'' under simple shear. We find that, instead of deforming uniformly, bacterial ``superfluids'' develop multiple shear bands, i.e., regions with different shear deformations. We construct a simple model that quantitatively describes the shape of the shear-banding structure and reveals important physical properties of collective bacterial motions. Our study sheds light on complex interactions between swimming microorganisms and ambient fluid flows, crucial for the survival of microorganisms in nature and the manipulation of bacterial suspensions in engineering settings.}

\authorcontributions{D.S. and X.C. conceived the project. S.G., D.S. and X.C. conducted the experiments and analyzed the data with help from Y.P. X.X. and X.C. developed the model. S.G., X.X. and X.C. wrote the manuscript with all authors contributed to the final version.}
\authordeclaration{The authors declare no conflict of interest.}
\equalauthors{\textsuperscript{1}Current address: Department of Mechanical Engineering, Indian Institute of Technology, Ropar, Nangal Road, Rupnagar, Punjab 140001, India}
\correspondingauthor{\textsuperscript{2}To whom correspondence should be addressed. E-mail: xinliang@csrc.ac.cn or xcheng@umn.edu}


\begin{abstract}
Bacterial suspensions---a premier example of active fluids---show an unusual response to shear stresses. Instead of increasing the viscosity of the suspending fluid, the emergent collective motions of swimming bacteria can turn a suspension into a ``superfluid'' with zero apparent viscosity. Although the existence of active ``superfluids'' has been demonstrated in bulk rheological measurements, the microscopic origin and dynamics of such an exotic phase have not been experimentally probed. Here, using high-speed confocal rheometry, we study the dynamics of concentrated bacterial suspensions under simple planar shear. We find that bacterial ``superfluids'' under shear exhibit unusual symmetric shear bands, defying the conventional wisdom on shear-banding of complex fluids, where the formation of steady shear bands necessarily breaks the symmetry of unsheared samples. We propose a simple hydrodynamic model based on the local stress balance and the ergodic sampling of nonequilibrium shear configurations, which quantitatively describes the observed symmetric shear-banding structure. The model also successfully predicts various interesting features of swarming vortices in stationary bacterial suspensions. Our study provides new insights into the physical properties of collective swarming in active fluids and illustrates their profound influences on transport processes.
\end{abstract}

\dates{This manuscript was compiled on \today}
\doi{\url{www.pnas.org/cgi/doi/10.1073/pnas.XXXXXXXXXX}}

\begin{document}

\verticaladjustment{-2pt}

\maketitle
\thispagestyle{firststyle}
\ifthenelse{\boolean{shortarticle}}{\ifthenelse{\boolean{singlecolumn}}{\abscontentformatted}{\abscontent}}{}

\dropcap{A}ctive fluids, suspensions of self-propelled particles, have attracted enormous research interests in recent years \cite{1,2,3,4,5}. With examples across biological and physical systems of widely different scales, active fluids exhibit many novel properties, such as the emergence of collective swarming \cite{6,7,8,9}, giant number fluctuations \cite{10,11} and enhanced diffusion of passive tracers \cite{12,13,14,15,16}. Among all these unusual features, the flow behavior of active fluids demonstrates the nonequilibrium nature of active systems in the most striking manner. Surprising phenomena including superfluid-like behaviors \cite{17} and spontaneous directional flows \cite{18,19} have been observed in active fluids.

Using a phenomenological model that couples hydrodynamic equations with active nematic order parameters, Hatwalne {\it et al.} first showed that pusher microswimmers such as {\it E. coli} can significantly lower the bulk viscosity of active suspensions, to such an extent that suspensions can have a lower viscosity than the suspending fluids \cite{20}. Based on a similar approach, Cates {\it et al.} further predicted that near the disorder-to-order transition to collective motions, a pusher active fluid can enter a ``superfluidic'' regime where its apparent shear viscosity vanishes \cite{21}. Later theory by Giomi {\it et al.} revealed even richer dynamics and predicted the existence of shear banding, yield stress, and ``superfluidity'' of active fluids \cite{22}. Unusual rheology of active fluids has also been studied based on the microhydrodynamics of microswimmers at low concentrations \cite{23,24,25,26,27}, swimming pressures \cite{28} and generalized Navier-Stokes equations \cite{29}. Experimentally, Sokolov {\it et al.} and Gachelin {\it et al.} showed the low viscosity of bacterial suspensions in thin films \cite{30,31}. Lopez {\it et al.} demonstrated the superfluid-like transition in concentrated {\it E. coli} suspensions using a rotational rheometer \cite{17}. Under channel confinements,  this ``superfluidic'' behavior displays as spontaneous directional flows \cite{18,19}. In comparison, puller swimmers such as swimming algae were shown to enhance, instead of suppressing, the viscosity of suspensions \cite{32}.  

\begin{figure*}
	\centerline{\includegraphics[width=0.85\linewidth]{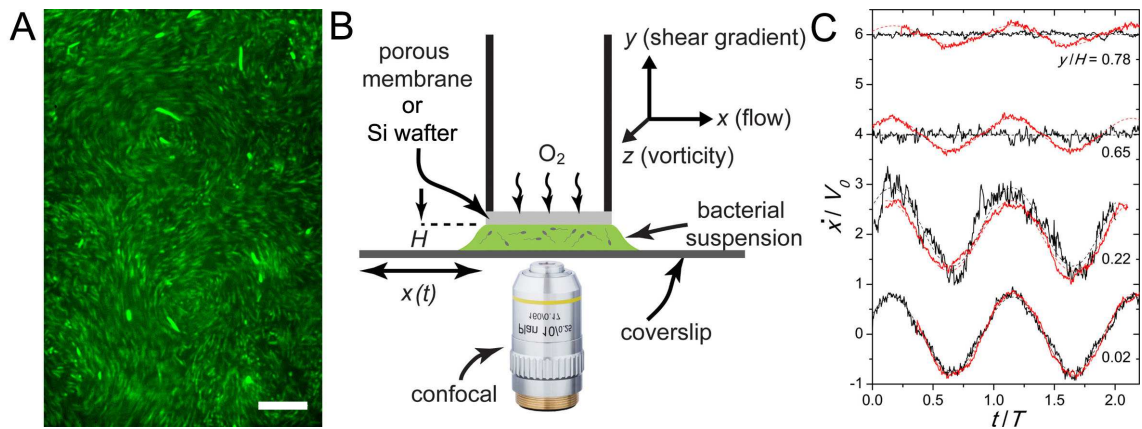}}
	\caption{Bacterial suspensions under planar oscillatory shear. (A) Bacterial swarming at a concentration $n = 80n_0$. The scale bar is 20 $\mu$m. The fluorescently tagged $E. coli$ serve as tracer particles for particle imaging velocimetry (PIV). (B) Schematic showing our custom shear cell. A Cartesian coordinate system is defined, where $x$, $y$ and $z$ are the flow, shear gradient and vorticity directions, respectively. (C) Temporal variation of mean suspension velocities $\dot{x}(t)$ at different heights, $y$, above the bottom plate. Red curves are for shear-rate amplitude $\dot{\gamma}_0 = 0.52$ s$^{-1}$. Black curves are for $\dot{\gamma}_0 = 0.21$ s$^{-1}$. Velocities are normalized by the imposed velocity amplitudes, $V_0$. Time $t$ is normalized by shear period $T = 1/f$. $y$ is normalized by gap thickness $H$. $\dot{x}(t)$ at different $y$ are shifted vertically for clarity. Dashed lines are sinusoidal fits. \label{Figure1}} 
\end{figure*}

Although the vanishing shear viscosity of active ``superfluids'' has been demonstrated in bulk rheology studies \cite{17}, the microscopic dynamics of such an exotic phase under simple shear flows have not be experimentally explored. The shear-banding structure---an important prediction of hydrodynamic theories \cite{21,22}---has not been verified. Here, using fast confocal rheometry, we study the dynamics of concentrated bacterial suspensions under planar oscillatory shear. We find that bacterial superfluids exhibit symmetric shear-banding flows with three shear bands. We systematically investigate the variation of the shear-banding structure with shear rates, bacterial concentrations and bacterial motility. Based on the existing hydrodynamic theories, we construct a simple phenomenological model that quantitatively describes the shape of the symmetric shear bands. The model also predicts several nontrivial properties of swarming vortices in stationary bacterial suspensions, including the linear relation between the kinetic energy and the enstrophy of suspension flows and the system-size dependence of the length and strength of swarming vortices. We conclude the paper by discussing the unique feature of the shear-banding flow of bacterial suspensions in comparison with conventional shear-banding complex fluids. Our study provides new insights into the collective swarming of active fluids and illustrates the unexpected consequence of collective swarming on momentum transports of active systems. Our results also help to understand complex interactions between bacteria and ambient shear flows encountered in many natural and engineering settings.


\section*{Results}

We use fluorescently tagged {\it Escherichia coli} K-12 strain (BW25113). The bacteria are suspended in a mobility buffer to a concentration $n$. We vary $n$ between $10n_0$ and $100n_0$ with $n_0 = 8 \times 10^8$ ml$^{-1}$ the concentration of bacteria in the stationary phase of growing. When $n \gtrsim 40n_0$, collective bacterial swarming can be observed (Fig.~\ref{Figure1}A and SI Appendix A).

We investigate the 3D fluid flow of {\it E. coli} suspensions under planar oscillatory shear. A suspension of 20 $\mu$l is confined between the two parallel plates of a custom shear cell with a constant spacing $H = 60$ $\mu$m unless other stated (Fig.~\ref{Figure1}B)\cite{33,34}. A circular top plate of radius 2.5 mm is stationary, whereas a much larger bottom plate driven by a piezoelectric actuator moves sinusoidally with $x(t) = A_0\sin(2\pi ft)$. The shear amplitude and frequency, $A_0$ and $f$, determine the amplitude of imposed shear rates, $\dot{\gamma}_0=V_0/H=2\pi fA_0/H$, where $V_0$ is the applied velocity amplitude. For most experiments, we vary $\dot{\gamma}_0$ by changing $A_0$ and fixing $f = 0.1$ Hz, although low shear frequencies for steady-state shear ranging from 0.025 to 0.3 Hz have also been tested (SI Appendix B). The bottom plate is made of a smooth glass coverslip, enabling us to image 3D suspension dynamics via an inverted confocal microscope. The top plate is made of either a smooth silicon wafer or a rough porous membrane that allows for the influx of oxygen (SI Appendix A). While the symmetric smooth shear boundary with the Si wafer eliminates the biased influence of the boundary on shear profiles, the porous membrane allows us to maintain high bacterial activities for $n >65n_0$ \cite{7}. Both shear boundaries yield qualitatively similar results.

\subsection*{Symmetric shear banding}   
\begin{figure}[t]
	\centerline{\includegraphics[width=0.8\linewidth]{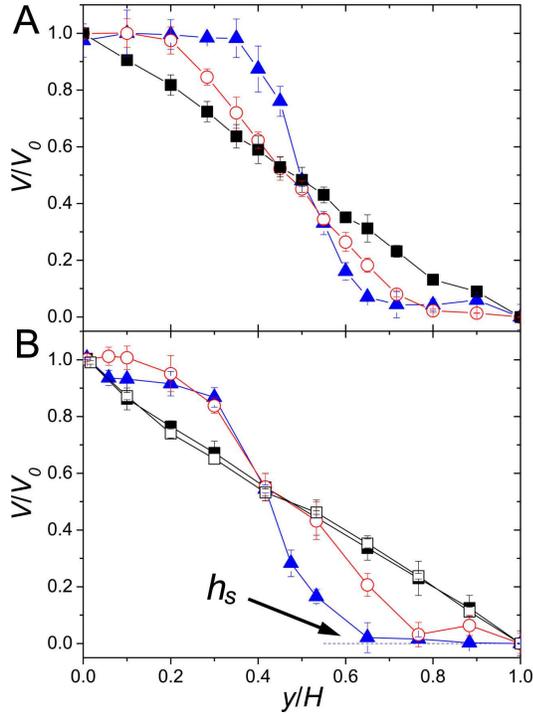}}
	\caption{Shear profiles of bacterial suspensions. (A) Normalized shear profiles at different shear rates. $V_0$ is the applied shear velocity amplitude. Bacterial concentration is fixed at $n = 50n_0$. The shear-rate amplitude $\dot{\gamma}_0 = 0.42$ s$^{-1}$ (black squares), 0.16 s$^{-1}$ (red circles) and 0.055 s$^{-1}$ (blue triangles). Si wafer is used as the top plate. (B) Normalized shear profiles at different bacterial concentrations. $\dot{\gamma}_0$ is fixed at $0.16$ s$^{-1}$. $n =10n_0$ (black squares), $40n_0$ (red circles) and $100n_0$ (blue triangles). To maintain bacterial motility at high $n$, a porous membrane is used as the top plate. The stop height, $h_s$, of the profile at $100n_0$ is indicated. Empty squares are for a suspension of immobile bacteria at $100n_0$. \label{Figure2}} 
\end{figure}

The average velocity of a concentrated bacterial suspension under shear at different heights $y$ above the bottom plate, $\dot{x}(y,t)$, is shown in Fig.~\ref{Figure1}C. Here, the average is taken along both the flow ($x$) and the vorticity ($z$) directions. $\dot{x}(y,t)$ is sinusoidal following $\dot{x}(y,t) = V(y)\cos(2\pi ft)$, where $V(y)$ is the velocity amplitude at $y$. A drastic difference in suspension dynamics can be identified between suspensions in the normal phase under strong shear and those in the ``superfluidic'' phase under weak shear \cite{17}. Under strong shear, $V(y)$ decreases linearly with $y$, similar to the response of dilute colloidal suspensions (Fig.~\ref{Figure2}A). However, under weak shear, interesting nonlinear shear profiles are observed. All the applied shear concentrates near the center of the suspensions. Near the top and bottom plates, local shear gradients are small and may even vanish, resulting in approximately symmetric shear profiles rarely seen in other complex fluids (Fig.~\ref{Figure2}A). A crossover from the linear to the nonlinear shear profiles is observed with decreasing $\dot{\gamma}_0$.    

The shape of shear profiles also depends on the strength of collective bacterial swarming. We vary the swarming strength by changing bacterial concentrations $n$ (SI Appendix Fig. S2) \cite{6}. At large $n$, bacteria show strong collective motions, leading to the nonlinear shear profiles at low $\dot{\gamma}_0$ (Fig.~\ref{Figure2}B). Below $40n_0$ where the collective swarming is weak, the shear profile appears to be linear even at low $\dot{\gamma}_0$. A similar crossover to the linear profile is also observed when bacterial swarming weakens due to the depletion of oxygen. A concentrated suspension of immobile bacteria shows a linear shear profile at all $\dot{\gamma}_0$ (Fig.~\ref{Figure2}B).  

\begin{figure}[t]
	\centerline{\includegraphics[width=0.78\linewidth]{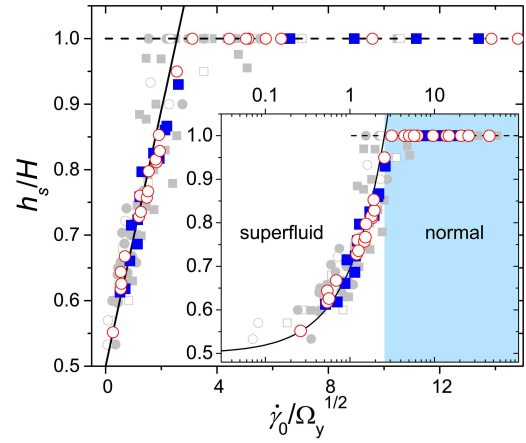}}
	\caption{Shape of shear profiles. The stop height, $h_s$, as a function of the dimensionless shear rate $\dot{\gamma}_0/\sqrt{\Omega_y}$. $h_s$ is normalized by the gap thickness $H$. $H = 30$ $\mu$m (squares) and 60 $\mu$m (circles). Color symbols are obtained with the symmetric shear boundary using Si wafer at $f = 0.1$ Hz. Gray symbols are obtained with the asymmetric shear boundary using the porous membrane. Solid gray symbols are for $f = 0.1$ Hz and empty gray symbols are for other shear frequencies between 0.025 Hz and 0.3 Hz. Inset shows the same data in a log-linear plot. The solid line is the theoretical prediction in the ``superfluidic'' phase and the dashed line is the prediction in the normal phase (Eq.~[\ref{h_s}]). \label{Figure3}} 
\end{figure}

The competition between the shear flow and the collective bacterial swarming dictates the microscopic suspension dynamics. The strength of shear flows is naturally quantified by the imposed shear rate amplitude, $\dot{\gamma}_0$. The strength of bacterial swarming can be quantified by the enstrophy of bulk stationary suspension flows without external shear, $\Omega_y \equiv \langle \omega_y^2/2 \rangle$ \cite{9}. Here, $\omega_y=\partial_x v_z-\partial_z v_x$ is the in-plane vorticity, where $v_x$ and $v_z$ are local suspension velocities along the flow and vorticity directions. The average $\langle \cdot \rangle$ is again taken over the flow--vorticity plane. We then construct a dimensionless shear rate $\dot{\gamma}_0/\sqrt{\Omega_y}$. To characterize the shape of shear profiles, we measure the stop height, $h_s$, above which the shear flow vanishes (Fig.~\ref{Figure2}B). $h_s$ is obtained experimentally by fitting shear profiles piecewise with three linear lines (SI Appendix Fig. S4). When plotting $h_s$ as a function of $\dot{\gamma}_0/\sqrt{\Omega_y}$, all our data at different imposed shear rates, bacterial activities and gap thicknesses collapse onto a master curve (Fig.~\ref{Figure3}). Above $\dot{\gamma}_0/\sqrt{\Omega_y} \approx 2$, the shear profiles are linear with $h_s = H$. At small $\dot{\gamma}_0/\sqrt{\Omega_y}$, $h_s$ increases linearly with $\dot{\gamma}_0/\sqrt{\Omega_y}$ and approaches $H/2$ in the zero shear limit.

\subsection*{Model} 
\begin{figure}[t]
	\centerline{\includegraphics[width=0.9\linewidth]{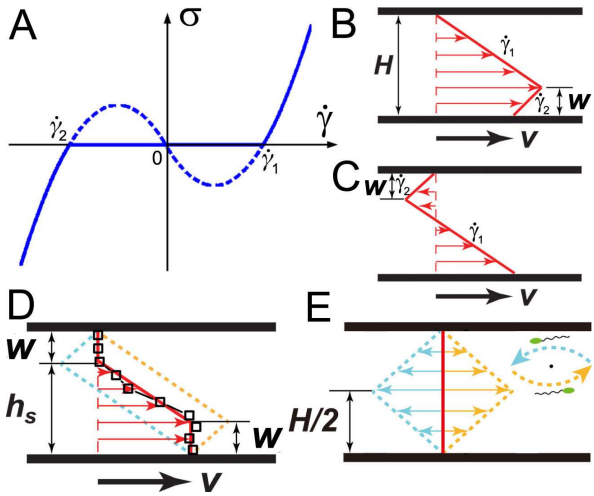}}
	\caption{Duality of shear configurations. (A) A schematic showing the constitutive relation of active fluids from hydrodynamic theories \cite{21,22}. The nonmonotonic trend predicts shear-banding flows with two shear bands of opposite shear rates, $\dot{\gamma}_1=\dot{\gamma}^*$ and $\dot{\gamma}_2=-\dot{\gamma}^*$. The corresponding shear profile are shown in (B) and (C). Red arrows indicate shear velocities at different heights. Gap thickness, $H$, and the width of the shear band with $\dot{\gamma}_2$, $w$, are indicated. (D) Symmetric shear profile (thick red line) resulting from the average of the two shear configurations in (B) and (C) (yellow and blue dashed lines). Symbols are the experimental shear profile at $n = 80n_0$ and $\dot{\gamma}_0$= 0.26 s$^{-1}$. The stop height, $h_s$, is indicated. (E) The duality of shear profiles at zero applied shear rate $\dot{\gamma}_0=0$. The mean flow is zero (thick red line), whereas the two shear-banding configurations (yellow and blue dashed lines) are symmetric with respect to the mean flow. Inset of (E): at given $y$, the two configurations moving along and against the shear flow complete a bacterial vortex in the $x$-$z$ plane. \label{Figure4}} 
	
\end{figure}

The existence of bacterial ``superfluids'' have been predicted by hydrodynamic theories of active fluids \cite{21,22}. These theories show that the constitutive equation of active fluids is nonmonotonic across zero (Fig.~\ref{Figure4}A). The mechanical instability induced by the negative slope of the constitutive relation then leads to a zero-stress ``superfluidic'' plateau \cite{35,36}. The instability also predicts a nonmonotonic shear profile with two shear bands of opposite shear rates (Fig.~\ref{Figure4}B). 

To understand the symmetric shear profiles in our experiments, we construct a simple phenomenological model based on the constitutive equation of the hydrodynamic theory \cite{21} (SI Appendix C). The local total shear stress, $\sigma_t$, can be divided into two parts, $\sigma_t=\sigma_s+\sigma_a$, where $\sigma_s= \eta \dot{\gamma}_{loc}$ is the local viscous shear stress with suspension viscosity $\eta$ and local shear rate $\dot{\gamma}_{loc}$. $\sigma_a=-|\sigma_a|\mathrm{sgn}(\dot{\gamma}_{loc})$ is the active stress originated from bacterial swimming \cite{22}, where sgn is the sign function. Here, we assume that the degree of local nematic ordering of bacteria is determined by steric and hydrodynamic interactions between bacteria, whereas the orientation of the nematic order is selected by the local shear flow. $|\sigma_a|$ is a function of bacterial concentrations and motility, but is insensitive to the magnitude of local shear rates \cite{21,22}. A shear-rate-dependent $|\sigma_a|$ based on detailed hydrodynamic theories does not change the predictions of our simple model (SI Appendix F). For simplicity, we also ignore the complex bacteria-boundary interaction, which may influence the average bacterial orientation near walls \cite{37}. Considering the bacteria-boundary interaction should not affect the key predictions of our model either (SI Appendix G).     

In the ``superfluidic'' phase, the stress balance, $\sigma_s+\sigma_a=0$, gives rise to two solutions, i.e., $\dot{\gamma}_{loc}=\dot{\gamma}^*$ and $\dot{\gamma}_{loc}=-\dot{\gamma}^*$, where $\dot{\gamma}^*\equiv|\sigma_a|/\eta$ is the characteristic shear rate of bacterial suspensions. To satisfy the no-slip boundary condition, we have the nonmonotonic shear-banding flow (Figs.~\ref{Figure4}B and C), where the width of the shear band with $-\dot{\gamma}^*$, $w$, follows (SI Appendix C)
\begin{equation}
\frac{w}{H}=\frac{1}{2}\left(1-\frac{\dot{\gamma}_0}{\dot{\gamma}^*} \right)=\frac{1}{2}\left(1-\frac{\dot{\gamma}_0}{C\sqrt{2\Omega_y}} \right) \label{h_d}
\end{equation}
Here, we replace $\dot{\gamma}^*$ by the experimental observable $\Omega_y$. In a stationary sample without external applied shear, bacterial swarming is solely driven by the active stress. Thus, the active stress balances the viscous stress, $|\sigma_a|= C\eta \omega_y= C\eta \sqrt{2\Omega_y}$, where $C$ is a proportionality constant close to one. Thus, $\dot{\gamma}^* = |\sigma_a|/\eta = C\sqrt{2\Omega_y}$. Since $w \geq 0$, $\dot{\gamma}_0 \leq \dot{\gamma}^*$, setting the necessary condition for ``superfluids''.

It should be emphasized that there are two and only two shear configurations with two shear bands satisfying the stress balance and the no-slip boundary condition, which are shown in Figs.~\ref{Figure4}B and C, respectively. Since both shear configurations satisfy the local stress balance, we hypothesize they emerge in a sheared sample ``ergodically'' with equal probability, an assumption that shall be tested {\it a posteriori}. The measured shear profile should then be an ``ensemble'' average of the two shear configurations. A possible physical interpretation is as follows: a single bacterial vortex normal to the flow-vorticity plane extending across the two shear plates (see Fig.~\ref{Figure1}A) can be viewed as composed of the two shear configurations (Fig.~\ref{Figure4}E inset). The half of the vortex moving along the shear direction represents the configuration of Fig.~\ref{Figure4}B, whereas the other half moving against shear gives the configuration in Fig.~\ref{Figure4}C. Thus, the ensemble average is achieved experimentally through a spatiotemporal average over multiple swarming vortices. Vortices have a characteristic diameter $\sim 60$ $\mu$m when $H = 60$ $\mu$m (Fig.~\ref{Figure1}A) and a life time of a few seconds \cite{7,29}, whereas the spatial and temporal scales of our experiments are 180 $\mu$m and 40 s, respectively. 

The ensemble average of the two shear configurations naturally leads to a symmetric shear profile (Fig.~\ref{Figure4}D), consistent with our observations. Using Eq.~[\ref{h_d}] and a simple geometric relation $h_s + w = H$, we have
\begin{equation}
\frac{h_s}{H} = \begin{cases}
\frac{1}{2}\left(1+\frac{\dot{\gamma}_0}{C\sqrt{2\Omega_y}}\right) & \text{if } \dot{\gamma}_0/\sqrt{\Omega_y} \leq C\sqrt{2}, \\
1 &  \text{if } \dot{\gamma}_0/\sqrt{\Omega_y} > C\sqrt{2}
\end{cases} \label{h_s},
\end{equation}
which successfully predicts the linear relation between $h_s$ and $\dot{\gamma}_0/\sqrt{\Omega_y}$ in the ``superfluidic'' phase (Fig.~\ref{Figure3}). A quantitative fitting of experimental data shows $C = 1.6 \pm 0.4$ on the order of one as expected. Notice that the two-band shear configurations in Figs.~\ref{Figure4}B and C are achieved in our experiments via the 1D confinement imposed by our shear cell along the shear gradient direction. At sufficient large $H$, three or more shear bands may emerge, which have infinite possible shear configurations satisfying the stress balance and the no-slip boundary condition. The ergodic assumption would then lead to featureless linear shear profiles (SI Appendix D). Our experiments are different from earlier studies on bacterial suspensions under channel confinement, which constrains bacterial swarming along both the shear gradient and vorticity directions. Such a confinement suppresses the instability that induces swarming vortices \cite{4}. As a result, suspensions develop directional flows and break the hypothesized ``ergodicity'' \cite{18}.

\begin{figure}[t]
	\centerline{\includegraphics[width=1\linewidth]{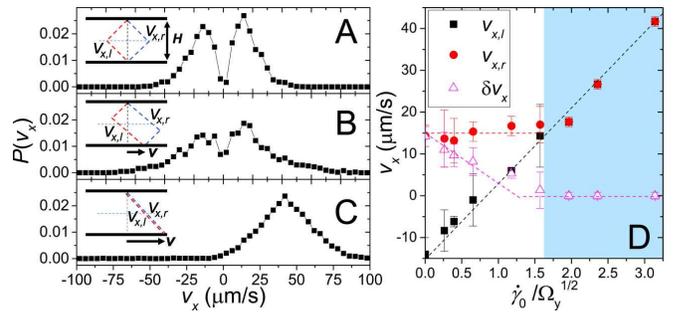}}
	\caption{Probability distribution function of local velocities along the flow direction, $v_x$, at different shear rates, $P(v_x)$. (A) $\dot{\gamma}_0/\sqrt{\Omega_y} = 0$, (B) $\dot{\gamma}_0/\sqrt{\Omega_y} = 0.24$ and (C) $\dot{\gamma}_0/\sqrt{\Omega_y} = 2.88$. Local velocities are measured when the average shear velocity reaches maximal in each shear cycle. PIV box size is chosen at $R$, where $R$ is the characteristic radius of swarming vorticies. $n= 80n_0$ and $H =60$ $\mu$m. Insets show schematically the corresponding shear profiles. The thick dashed lines (red and blue) indicate the two shear configurations. The thin horizontal dashed line indicates the position of our imaging plane. The intersections give two discrete velocities, $v_{x,l}$ and $v_{x,r}$, corresponding to the two peaks of $P(v_x)$. (D) The two peaks of $P(v_x)$, $v_{x,l}$ (black squares) and $v_{x,r}$ (red disks), and velocity variance, $\delta v_x$ (magenta triangles), as a function of shear rate, $\dot{\gamma}_0/\sqrt{\Omega_y}$. Dashed lines show the model predictions.  \label{Figure5}} 
	
\end{figure}

The model incorporates an unique feature, i.e., a dynamic alternation between the two shear configurations around the mean shear profile (Fig.~\ref{Figure4}D). To verify the hypothesis, we measure the probability distribution function of local velocities at the center of the shear cell, $P(v_x)$, at different shear rates (Fig.~\ref{Figure5}). At zero and low shear rates (Figs.~\ref{Figure5}A and B), bimodal distributions with two distinct peaks can be identified. The peaks correspond to the velocities of the two discrete shear profiles at $y = H/2$, $v_{x,l}$ and $v_{x,r}$  (Figs.~\ref{Figure5}A and B insets). The finite width of the distributions arises presumably from the variation of individual bacterial mobility, an effect that is not included in our model. The areas underneath the two peaks are approximately the same with difference less than $5\%$ at $\dot{\gamma}_0=0$, supporting our ergodic assumption. At high shear rates in the normal phase, $P(v_x)$ becomes unimodal (Fig.~\ref{Figure5}C), indicating the emergence of a single linear profile (Fig.~\ref{Figure5}C inset). Our model predicts that the left peak of $P(v_x)$, $v_{x,l}$, increases linearly with $\dot{\gamma}_0$ in both phases, whereas the right peak of $P(v_x)$, $v_{x,r}$, is constant in the ``superfluidic'' phase and merges with $v_{x,l}$ in the normal phase. The variance of velocity $\delta v_x$ from the model follows (SI Appendix E)    
\begin{equation}
\delta v_x = \sqrt{\langle v_x^2 \rangle - \langle v_x \rangle^2} = \frac{v_{x,r}-v_{x,l}}{2} = \frac{H}{2}\left( \sqrt{2\Omega_y}-\dot{\gamma}_0 \right) \label{variation}
\end{equation}
in the ``superfluidic'' phase and becomes zero in the normal phase. Our experiments quantitatively agree with all these predictions (Fig.~\ref{Figure5}D). Direct measurements on instantaneous shear profiles at local scales are certainly needed to finally verify the ergodic assumption of our model, which is constructed to rationalize the 3D experimental results using simple steady-state 1D shear profiles (SI Appendix C).    

\subsection*{Swarming vortices in stationary bacterial suspensions}
\begin{figure}[t]
	\centerline{\includegraphics[width=1\linewidth]{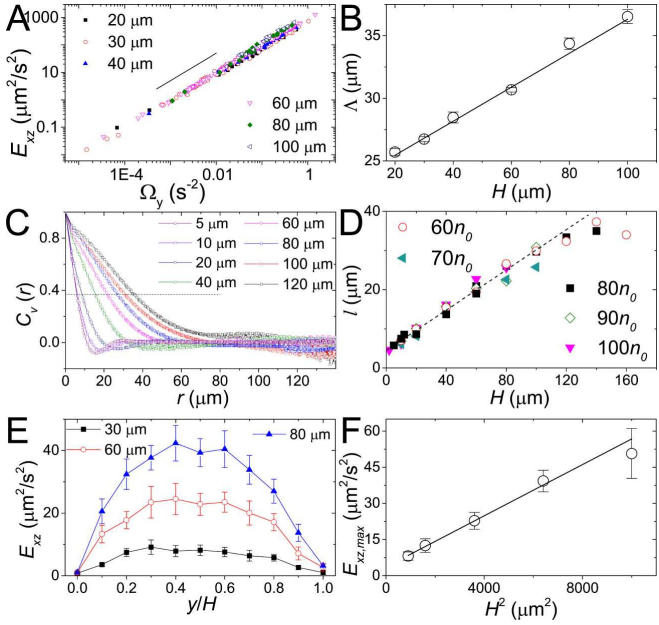}}
	\caption{Properties of bacterial swarming in stationary samples. (A) Kinetic energy, $E_{xz}$, versus enstrophy of suspension flows, $\Omega_y$. The gap size $H$ is indicated in the plot. Flows are measured at the midplane $y=H/2$. The solid line indicates the linear relation $E_{xz} \sim \Omega_y$. (B) $\Lambda$ extracted from the slope of $E_{xz}(\Omega_y)$ versus $H$. The solid linear is a linear fit. (C) Velocity spatial correlations. The horizontal dashed line is $e^{-1}$. $H$ is indicated. (D) Correlation length, $l$, as a function of $H$. Bacterial concentrations, $n$, are indicated. The dashed line indicates the linear relation. (E) $E_{xz}$ as a function of the height $y$ at three different $H$. $n=64n_0$. (F) The maximum $E_{xz}$ at $y=H/2$ versus $H^2$. The solid line is a linear fit. \label{Figure6}} 
	
\end{figure}

The minimal model also predicts several nontrivial properties of swarming vortices in stationary bacterial suspensions without shear. First, from Eq.~[\ref{variation}], when $\dot{\gamma}_0=0$, $\delta v_x^2 = \Lambda^2 \Omega_y$, where $\Lambda^2 \equiv H^2/2$. Since without shear $\langle v_x \rangle = 0$, $\delta v_x^2 = \langle v_x^2 \rangle$. The kinetic energy of suspension flows $E_{xz} = \langle v_x^2 \rangle = \delta v_x^2 = \Lambda^2 \Omega_y$. Thus, the model predicts that the kinetic energy of a bacterial swarming flow is linearly proportional to the enstrophy of the flow. The square root of the slope, $\Lambda$, is proportional to the gap size of the system. Although the linear relation between $E_{xz}$ and $\Omega_y$ has been reported in experiments on thin bacterial films and in simulations using generalized Navier-Stokes equations \cite{9}, there still lacks a simple physical explanation of its origin. Our simple model shows that such a linear relation arises from the alternation of self-organized shear profiles in unsheared samples dictated by the local stress balance. To verify the model, we measure $E_{xz}$ and $\Omega_y$ of stationary bacterial suspensions. At a fixed $H$, $E_{xz}$ indeed increases linearly with $\Omega_y$ for different bacterial motility (Fig.~\ref{Figure6}A). More importantly, we measure $E_{xz}(\Omega_y)$ at different gap sizes and extract $\Lambda$ from the slope of the linear relations. $\Lambda$ as a function of $H$ shows a clear linear trend (Fig.~\ref{Figure6}B), agreeing with the model, although the slope of $\Lambda(H)$ is smaller than the predicted value.   

Previous studies implied that $\Lambda$ is associated with the length scale of swarming vortices \cite{9}. Since $\Lambda$ changes linearly with $H$ (Fig.~\ref{Figure6}B), we hypothesize that the size of swarming vortices should also change linearly with the gap size of the system. To test the hypothesis, we measure the velocity--velocity spatial correlation (Fig.~\ref{Figure6}C)
\begin{equation}
C_{v}(r) = \frac{\iint \mathrm{d}\vec{r}_i \mathrm{d}\vec{r}_j (\vec{v}(\vec{r}_i)\cdot\vec{v}(\vec{r}_j) ) \delta(r_{ij}-r) }{\int \mathrm{d}\vec{r}_i (\vec{v}(\vec{r}_i) \cdot \vec{v}(\vec{r}_i))},
\end{equation}
where the local suspension velocity $\vec{v}$ is measured at the midplane $y=H/2$ and $r_{ij}= |\vec{r}_i-\vec{r}_j|$. The correlation length $l$ of swarming vortices is extracted from the location where $C_v$ decreases to $1/e$. $l$ as a function of $H$ is shown in Fig.~\ref{Figure6}D. A linear relation is observed when $H < 120$ $\mu$m. Our results are consistent with previous published data using different experimental setups. In thin chambers with height $< 10$ $\mu$m, the vortex size is of the order of $5-10$ $\mu$m \cite{8}, whereas in chambers of height $\sim 80$ $\mu$m the vortex size increases to $\sim 50$ $\mu$m \cite{9}. At even larger $H$, $l$ shows a trend for saturation. Although the working distance of the confocal microscope prevents us from imaging samples with very large $H$, a large swarming vortex with strong nematic order is known to be unstable for pusher suspensions \cite{1,3,4,20}.

Lastly, the two shear configurations are symmetric without shear, leading to zero mean velocity (Fig.~\ref{Figure4}E). $E_{xz} = \delta v_x^2$ shows a non-monotonic trend with $y$, where $E_{xz}$ reaches a maximum, $E_{xz,\max}$, at the center of the cell and approaches to zero at the top and bottom walls. Our experiments confirm the nonmonotonic trend of $E_{xz}(y)$ (Fig.~\ref{Figure6}E). Since the local shear gradient $\dot{\gamma}^*$ is independent of the gap size $H$, as we increase $H$, the velocity fluctuation $\delta v_x$ should increase linearly with $H$. Thus, $E_{xz,\max}$ should increases as $H^2$. Our experiments quantitatively agree with this prediction (Fig.~\ref{Figure6}F). Thus, in addition to the length scale of swarming vortices, the model also successfully predicts the dependence of their strength, characterized by $E_{xz}$, on the system size.            

\begin{figure}[t]
	\centerline{\includegraphics[width=0.95\linewidth]{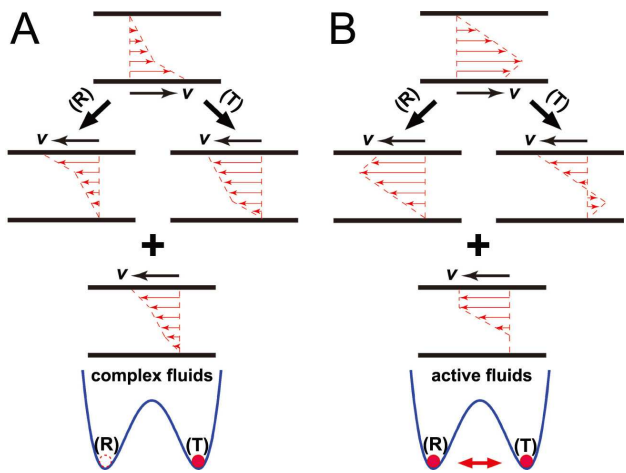}}
	\caption{Comparison of shear banding in complex and active fluids. (A) Shear banding in conventional complex fluids. The shear banding flow breaks the symmetry of unsheared samples, which can be seen from the difference in the shape of shear profiles after two physical operations: (i) a rotational operation (R), where the system is rotated counter-clockwise by $\pi$, and (ii) a translational operation (T), where the lab frame is transformed into a moving frame of a linear velocity $-V$. Although the boundary conditions of the systems after the two operations are the same, the resulting shear profiles are different. Thus, the sheared sample before the operations cannot simultaneously satisfy the translational and rotational symmetry of the unsheared sample. The ensemble average of the two symmetry-broken shear configurations is approximately linear, restoring the original symmetry of the unsheared sample. A sheared complex fluid chooses one of the two symmetry-broken configurations depending on initial and/or boundary conditions. The symmetry-broken process is illustrated schematically by the location of a red disk in a split-bottom potential, in analogy to the spontaneous symmetry breaking in equilibrium phase transitions. The valleys (R) and (T) indicate the two possible symmetry-broken shear-banding configurations. (B) Shear banding in active fluids. The ensemble-averaged shear profile from the two symmetry-broken shear-banding configurations is symmetric and nonlinear. A sheared active fluid samples both symmetry-broken configurations and preserves the symmetry of the unsheared fluid. \label{Figure7}} 
\end{figure}

\subsection*{Comparison with other shear-banding complex fluids}
Our study on 3D suspension dynamics shows that bacterial ``superfluids'' arise from the balance of local viscous and active stresses. Moreover, the duality of shear-banding configurations reveals a remarkable feature of active fluids, different from the shear-banding behavior of equilibrium complex fluids such as worm-like micelle solutions \cite{38}, colloidal suspensions \cite{39} and entangled polymeric fluids \cite{40}. Shear rates in these complex fluids are invariably positive \cite{35,36}. The formation of shear bands necessarily breaks the translational and rotational symmetry of the unsheared samples (Fig.~\ref{Figure7}A). Although the lost symmetry can be restored theoretically when all allowed shear-banding configurations are averaged, a shear-banding complex fluid invariantly selects one of the symmetry-broken configurations in the steady state (Fig.~\ref{Figure7}A). The choice of the specific configuration depends on initial and/or boundary conditions, a process analogous to the spontaneous symmetry breaking in phase transitions. In contrast, a sheared active fluid, instead of being trapped into one of the symmetry-broken configurations, samples all allowed shear-banding configurations (Fig.~\ref{Figure7}B), which leads to a symmetric yet nonlinear shear profile preserving the original symmetry of the unsheared sample. Although an active fluid is intrinsically out of equilibrium, it appears to be more ``ergodic'' due to its collective motions.

\section*{Conclusions}
We investigated the dynamics of concentrated bacterial suspensions under simple oscillatory shear using fast confocal rheometry. We observed unusual symmetric shear-banding flows in the ``superfluidic'' phase of bacterial suspensions, rarely seen in conventional complex fluids. A minimal phenomenological model was constructed based on the detailed stress balance and the ergodic sampling of different shear configurations, which quantitatively describes the variation of the shear-banding structure with applied shear rates and bacterial activity.  Such a simple model also successfully predicts various non-trivial physical properties of collective swarming in stationary bacterial suspensions. Particularly, it explains the linear relation between the kinetic energy and the enstrophy of suspension flows and shows the dependence of the length and strength of swarming vortices on the system size. Our study provides new insights into the emergent collective behavior of active fluids and the resulting transport properties. It illustrates the unusual rheological response of bacterial suspensions induced by the complex interaction between bacteria and ambient shear flows, which is frequently encountered in natural, biomedical and biochemical engineering settings.

\showmatmethods{} 

\acknow{We thank  K. Dorfman, Y.-S. Tai and K. Zhang for helps with bacterial culturing and J. Brady and Z. Dogic for discussions. The research is supported by DARPA YFA (No. D16AP00120), the Packard Foundation and NSF-CBET (No. 1702352). X. X. acknowledge support from National Natural Science Foundation of China (No. 11575020) and (No. U1530401).}

\showacknow{} 



\end{document}